\documentclass[a4paper,10pt]{article}

\title{Quantum Field Theory in de Sitter Spacetime }
\author{Ashaq Hussain Sofi$^1$,   Muhammad Ashraf Shah$^1$, \\
 Marlina Rosalinda Sibuea$^2$, Shabir Ahmad Akhoon$^1$,\\
Bilal Nisar Khanday$^3$, Sajad Ul Majeed$^3$, 
\\
Asloob Ahmad Rather$^3$,  Ishaq Nahvi$^4$ \\
$^1$ Department of Physics, National Institute of Technology, \\ Srinagar, Kashmir-190006,  India
\\ $^2$ Dexa Medica, Titan Center 3rd Floor, \\ 
Jalan Boulevard Bintaro Block B7/B1 No. 05 \\
Bintaro Jaya Sector 7,  Tangerang 15224, Indonesia\\
$^3$ Department of Physics, University of Kashmir, \\ Srinagar, Kashmir-190006,  India\\
$^4$Bemina Degree Collage, Srinagar, Kashmir-190018, India 
}

\begin{document}

\maketitle

\begin{abstract}
In this paper we will analyse  quantum field theory on de Sitter spacetime. 
We will  analyse a general scalar and vector field theory on de Sitter spacetime. 
This is done by first calculating these propagators on four-Sphere and then analytically continuing  
it to de Sitter spacetime. 
\end{abstract}

\section{Introduction}
 Gauge theory is one of the most fundamental theories of nature  \cite{a}-\cite{b}.
To quantize a gauge theory we need to fix the gauge gauge degrees of freedom \cite{e}-\cite{smu}. 
Gauge theory can also be studied in curved spacetime
\cite{7a}-\cite{wd1}. 
De Sitter spacetime has a positive cosmological constant that sets the rate of expansion.
The cosmological constant $\lambda$  is related to Hubble's constant $H$ of de Sitter spacetime as follows:
$ \lambda = 3 H^2
$. Hubble's constant is the inverse of the radius of  de Sitter spacetime. 
So  de Sitter spacetime is an expanding or contracting three-spheres with the radius $R$,  which is given by
$ R = \frac{1}{H}.
$ Thus de Sitter spacetime can be viewed as  a hyper-surface in five dimensional Minkowski spacetime. 
If the coordinates in the five dimensional Minkowski spacetime are  $X, Y, Z, W, T $  and the metric is given by
$  ds^2 = -dT^2 + dW^2 +dX^2 + dY^2 + dZ^2,
$ then de Sitter is the hyper-surface determined by the following equation
$  X^2 + Y^2 +Z^2 + W^2 - T^2 = R^2.
$ Now the metric in de Sitter spacetime can  be obtained by substituting the following in the Minkowski metric
\cite{ql1}-\cite{ql0}.
$  T =R \sinh t,
  W = R \cosh t \cos \chi ,
  X = R \cosh t \sin \chi \sin \theta \cos \phi ,
  Y = R \cosh t \sin \chi \sin \theta \sin\phi ,
  Z = R \cosh t \sin\chi \cos \theta.
$ Thus  the metric in de Sitter spacetime is  as follows:
 \begin{eqnarray}
  ds^2 &=& R^2 [-dt^2 + \cosh^2 t ds_3^2 ]\nonumber \\ &=& \frac{1}{H^2} [-dt^2 + \cosh^2 t ds_3^2 ].
 \end{eqnarray}
Here $ds_3^2$ is the metric on three-sphere. If we Euclideanize,  we get the metric four-sphere as follows:
\begin{equation}
 ds_4^2 = \frac{1}{H^2}[d\Phi^2 + \sin \Phi^2 ds_3^2],
\end{equation}
where $\Phi$ is given by
$  \Phi = \frac{\pi}{2} -it.
$ We can also introduce a new variable $\tau$ given by
$  \tan \tau = \sinh t.
$ Then the metric will be given by 
\begin{equation}
 ds^2 = \frac{1}{H^2\cos^2 \tau}[-d\tau^2 + ds_3^2].
\end{equation}
 where we have
$   T = R \tan\tau, 
  W = R \sec \tau \cos \chi , 
  X = R \sec\tau\sin \chi \sin \theta \cos \phi , 
  Y = R \sec\tau \sin \chi \sin \theta \sin\phi , 
  Z = R \sec \tau \sin\chi \cos \theta.
$ We could also write the de Sitter metric as follows:
\begin{equation}
 ds^2 = -dt^2 + \exp(2H t) (dx^2+dy^2+dz^2).
\end{equation}
where
$   t = R \log(W+T), 
  x = R \frac{X}{W+T},  
  y = R \frac{Y}{W+T}, 
  z= R \frac{Z}{W+T}. 
$ 

It may be noted that  
 the dimming of distant SN Ia apparent luminosity 
 has been mostly ascribed to the influence of 
 dark energy \cite{abcf1}-\cite{abcf2}. This has shown that our universe may approach 
de Sitter spacetime asymptotically.  
 In fact, the influences of inhomogeneities can lead to a departure of the observed 
 universe from an Einstein-de Sitter model \cite{asymp}. 
 It has also been demonstrated that  all the vacuum solutions of Einstein field equation with 
 a positive cosmological constant are models for de Sitter 
 spacetime \cite{ds}.
  The effects of noncommutativity of 
  spacetime geometry on the thermodynamical properties of the de Sitter horizon
  has also been studied \cite{ds1}.
 The  noncommutativity of spacetime  resulted in a modifications of the 
  temperature, entropy and vacuum energy of the de Sitter spacetime. 
  These modifications were  of the order of the Planck scale. Thus, it was 
  suggested that the size of the noncommutative parameter should be close to that of the Planck.
It may be noted that   the Weinberg-Salam model can be interpreted in terms of a gravity
    theory with the modulus of Higgs field as dilaton and the de Sitter space as the ground state
    \cite{ds2}. The gauge symmetry breaking in the Weinberg-Salam model 
can be implemented by a  change of variables and without any explicit gauge fixing. This  
 change of variables can entails the concept of supercurrent 
 which has been widely employed in the study of superconductivity. 
 In fact, it also introduces a separation between the isospin and the hypercharge,
 suggesting that the new variables describe a strongly coupled regime of the 
 electroweak theory. Thus,   the description
 of various embedded topological defects in terms of these variables have been also
 studied.
 The de Sitter vacua have also been studied in   supergravity and Calabi-Yau string models  \cite{ds4}.
In fact, a general analysis  on the possibility of obtaining 
metastable vacua with spontaneously broken  supersymmetry and non-negative cosmological 
constant in the moduli sector of string models has also been performed.
The condition under which the scalar partners of the Goldstino are non-tachyonic,
which depends only on the Kahler potential have thus been studied.
This condition has been found to be  necessary and sufficient, 
in the sense that all of the other scalar fields 
can be given arbitrarily large positive square masses if 
the superpotential is suitably tuned.
Both heterotic and orientifold string compactifications in the large-volume 
limit and show that the no-scale property shared by these models severely
restricts the allowed values for the sGoldstino masses in the superpotential parameter space.
Thus, a positive mass term has been  achieved only for certain types of compactifications and 
specific Goldstino directions.  
It may also be noted that    the presence of a
horizon in de Sitter space may be considered in conflict
with the idea that a horizon emits Hawking radiation in
an underlaying decaying process. The same discussion for
negative cosmological constants is solved by the presence,
or not, of Killing spinors because of their connection with
the definition of a BPS state \cite{ds5}. For a positive cosmological constant such a connection can not be established,
although Killing spinors indeed exist, because the de Sitter group does not have a supersymmetric extension.
Therefore, in order to understand the
role of the de Sitter as a ground state, 
 the thermodynamics of black holes with positive
cosmological constant is studdied \cite{ds6}. 
It may also be noted that 
quantum field theories on de Sitter spacetime with global gauge symmetry have been 
deformed using the joint action of the internal symmetry group and a one-parameter group of boosts
\cite{ds7}.
The resulting theory turned out to be wedge-local 
and non-isomorphic to the initial one for a class of theories, including the free charged Dirac field.
The properties of deformed models coming from inclusions of CAR-algebras were thus analysed. 
So, it turns out that the quantum field theory on de Sitter spacetime is an important subject. 
It may also be noted that de Sitter spacetime can be obtained from an analytical continuation
of a four sphere. Thus, quantum field theory can be obtained by first solving 
an equation using spherical harmonics and then analytical continuation of the solution thus obtained. 
This is what will be done in this paper. Thus, we will first solve the scalar and vector field 
equations on four sphere and then analytically continue the solution to de Sitter spacetime.

 \section{Formalism}
 In this section, we will write the expression for the Riemann curvature, 
 Ricci tensor and scalar curvature in de Sitter spacetime.
 In de Sitter spacetime the Riemann curvature $R^a_{bcd}$ is given by \cite{ql1}-\cite{ql0}, 
 \begin{equation}
  R^a_{bcd} = H^2 [\delta^a_c g_{bd} - \delta^a_d g_{bc} ].
 \end{equation}
Then it is straight forward to calculate $R_{bc}$ and it is given by
\begin{equation}
 R_{ab} = R^c_{acb} = 3H^2 g_{ab}.
\end{equation}
 Thus the scalar curvature $R$ is given by
 \begin{equation}
  R = g^{ab}R_{ab} = 12 H^2.
 \end{equation}
 This concludes our discussion of de Sitter spacetime.

Let $x$ and $x'$ be two spacelike separated points and 
let $\mu(x,x')$ be the geodesic distance between them. 
One defines a new variable $z$ on $S^4$ such that \cite{ql2}-\cite{ql3}
\begin{equation}
 z = \cos^2 \left(\frac{\mu}{2} \right).
\end{equation}
 One also defines the unit tangent vectors $n_a$ at $x$ and $n_{a'}$ at $x'$ along the geodesic between these two points as follows:
 \begin{equation}
  n_a = \nabla_a \mu (x, x')
 \end{equation}
 and
 \begin{equation}
  n_{a'} = \nabla_{a'} \mu (x, x').
 \end{equation}
In addition one defines a parallel propagator $g_{aa'}$ such that if  $A^a$ is a vector at $x$, then $A^{a'}$ is the vector at $x'$ obtained by parallelly transporting $A^a$ along the geodesic.
\begin{equation}
 A^{a'} = g^{a'}_a A^a.
\end{equation}
 Now as the unit tangents at $x$ and $x'$ point away from each other we have
 \begin{equation}
  g^{a}_{a'} n_a = - n_{a'}.
 \end{equation}
One also writes the metric at $x$ and $x'$ as $g_{ab}$ and $g_{a'b'}$ respectively.
Now any maximally symmetric bi-tensor can be expressed as a linear combination of  $g_{ab}$, $g_{a'b'}$, $n_a$, $n_{a'}$ and $g_{ab'}$ with the coefficient of each term depending only on $z$.
For example a bi-vector $Q_{ab'}$ may be expressed as
\begin{equation}
 Q_{ab'} = \alpha(z)g_{ab'} + \beta(z) n_a n_{b'}.
\end{equation}

The associated Legendre function $P^{-\mu}_{\nu}(x)$ is given by \cite{ql4}-\cite{ql5}
\begin{equation}
P^{-\mu}_{\nu}(x) = \frac{1}{\Gamma(1+\mu)}\left(\frac{1-x}{1+x}\right)^{\frac{\mu}{2}} F(-\nu, \nu+1, \mu+1, \frac{1-x}{2}), 
\end{equation}
where  $\Gamma(1+\mu) $ is the Gamma function.  
Furthermore,   the hypergeometric function
$F(\nu, \nu+1, \mu+1, \frac{1-x}{2})$ is the hypergeometric function 
The hypergeometric function  $F(a,b,c,x)$ can be written as 
 \begin{equation}
  F(a,b,c,x) = 1 + \frac{ab}{c}x + \frac{a(a+1)b(b+1)}{2! c(c+1)} x^2 + \cdots.
 \end{equation}
We also need to define operators that will lower or raise  $\nu$. 
The raising operator  is defined to be
\begin{equation}
 \left(  (1-x^2)\frac{d}{dx} -(\nu+1) x\right)P^{-\mu}_{\nu}(x) = -(\nu+\mu+1)P_{\nu+1}^{-\mu}(x).
\end{equation}
 and the lowering operator is defined to be 
\begin{equation}
 \left(  (1-x^2)\frac{d}{dx} +\nu x\right)P^{-\mu}_{\nu}(x) = (\nu-\mu)P_{\nu-1}^{-\mu}(x).
\end{equation}
We now define  $D_m$ as, 
\begin{equation}
 D_m = \frac{d}{d\chi} + m\cot\chi.
\end{equation}
Now we have 
\begin{eqnarray}
  \left[ \frac{d}{d\chi} + m\cot\chi \right] (\sin\chi)^n f(\chi) = && \nonumber \\
  \sin^n\chi \left[ \frac{d}{d\chi} + (m+n)\cot\chi \right] f(\chi),&&
\end{eqnarray}
and so we can write 
\begin{equation}
 D_m \sin^n \chi f(\chi) = \sin^n \chi D_{m+n} f(\chi).
\end{equation}
So, finally we have
\begin{equation}
 -\sin \chi D_n = \left[ (1- \cos\chi^2)\frac{d}{d\cos\chi} - n \cos\chi \right]. 
\end{equation}

\section{Scalar Field Theory on de Sitter Spacetime}
Now we will give a method for 
constructing scalar spherical harmonics on $S^n$ from scalar spherical harmonics on $S^{n-1}$
\cite{ql6}-\cite{ql7}.
Using this construction, we will construct scalar spherical harmonics on $S^4$. 
We first start with the metric on $S^n$, which is given by  
\begin{equation}
 ds^2 = d\chi^2 + \sin\chi^2 ds_{n-1}, 
\end{equation}
where the metric on $S^{n-1}$ is  $ds_{n-1}$.
Now let $m = (2-n)/2$ and also let $^n P^l_L$  be defined as
\begin{equation}
 {^n P^l_L} (\chi) =  c_n (\sin \chi)^m P^{-l+m}_{L-m}(\cos \chi),
\end{equation}
where $ c_n $ is a normalization  constant. It is given by
\begin{equation}
 c_n = \left[ \frac{(2L +n -1)(L+l+n-2)!}{2(L+l)!}\right]^{\frac{1}{2}}.
\end{equation}
The scalar spherical harmonics on $S^n$ can be defined as
\begin{equation}
 Y_{Llp...m}(\chi, \theta,\phi \cdots) = {^n P^l_{L}}(\chi) Y_{lp...m}(\theta, \phi \cdots),
\end{equation}
where $Y_{lpq....m}$ are the $n$ dimensional  scalar spherical harmonics 
and $Y_{ p q ....m}$ are the $n-1$ dimensional scalar spherical harmonics.
Now we define the LB operator in $n$ dimensions as $ - \nabla^2_n = - \nabla^a\nabla_a$, where 
$(a =1, 2....n) $. 
This LB operator $\nabla^2_{n-r}$  in $n-r$  dimensions  can also be defined to be 
the operator which satisfies, \cite{ql6}-\cite{ql7}
\begin{equation}
 -\nabla^2_{n-r} Y_{lpq...m} = l'(l'+n-r-1)Y_{lpq...m},
\end{equation}
where $l'$ is the index which comes in when we construct $n-r$ dimensional spherical harmonics.
Both these definitions are consistent with one another. 
Using this method, $n$ dimensional spherical harmonics can be constructed from 
 spherical harmonics on $S^1$.

 Now $Y_m$ is defined to be the 
 spherical harmonics on $S^1$. It is given by 
 \begin{equation}
  \frac{1}{\sqrt{2\pi}} \exp(im\phi).
 \end{equation}
Similarly, the spherical harmonics on $S^2$ is defined to be $Y_{pm}$. It is given by  
\begin{equation}
 Y_{pm} = c_2 {^2 P^m_p} Y_m.
\end{equation}
Furthermore, $Y_{lpm}$ is defined to be the spherical harmonics on $S^3$. It is given by 
\begin{equation}
  Y_{lpm} = c_3 {^3 P^p_l} Y_{pm}.
\end{equation}
 Finally, $Y_{Llpm}$ is defined to be the spherical harmonics on $S^4$. It is given by
 \begin{equation}
  Y_{Llpm} = c_4 {^4 P^l_L} Y_{lpm}.
 \end{equation}
 
The four dimensional  LB operator $\nabla^2_4$ is defined to be
be an operator that acts on $Y_{Llpm}$ and has $L(L+3) $ as its eigenvalue, 
\begin{equation}
 -\nabla_4^2 Y_{Llpm} = L(L+3) Y_{Llpm}.
\end{equation}
The three dimensional  LB operator $\nabla^2_3$ is defined to be
be an operator that acts on $Y_{Llpm}$ and has $l(l+2) $ as its eigenvalue, 
\begin{equation}
 -\nabla_3^2 Y_{Llpm} = l(l+2) Y_{Llpm}.
\end{equation}
The two dimensional  LB operator $\nabla^2_2$ is defined to be
be an operator that acts on $Y_{Llpm}$ and has $p(p+1) $ as its eigenvalue, 
\begin{equation}
- \nabla_2^2 Y_{Llpm} = p(p+1) Y_{Llpm}.
\end{equation}
The one dimensional  LB operator $\nabla^2_1$ is defined as
\begin{equation}
\nabla_1^2 =  \frac{\partial^2 }{\partial \phi^2}.
 \end{equation}
When it acts on $Y_{Llpm}$, the eigenvalue is $m^2$, 
\begin{equation}
 - \nabla_1^2 Y_{Llpm} = m^2 Y_{Llpm}.
\end{equation}
Furthermore,  these spherical harmonics  are normalized as, 
\begin{equation}
 \int d4x \sqrt{g} Y_{Llpm}Y^*_{L'l'p'm'} = \delta_{LL'}\delta_{ll'}\delta_{pp'}\delta_{mm'}, 
\end{equation}
where $g$ is the metric on $S^4$.

As we will only be dealing with 
the theory in four dimensions, 
we will write the LB operator on four dimensions as $-\nabla_4^2$ as $-\nabla^2$. 
So, we can write 
\begin{equation}
 [\nabla^2 + L(L+3) ] Y_{Llpm} = 0.
\end{equation}
Now, we start with a free massive scalar field  equation on de Sitter spacetime,
\begin{equation}
 [\nabla^2 - m^2 ] G (z) = 0.
\end{equation}
Thus after analytical continuation from $S^4$ to de Sitter,  we have 
\begin{equation}
 L(L+3) + m^2 = 0.
\end{equation}
Thus we get 
\begin{equation}
 L = a_{\pm} = -\frac{3}{2} \pm \sqrt{\frac{9}{4}- m^2}
\end{equation}
Using this the scalar field equation takes the form, 
\begin{equation}
 \left[z(1-z)\frac{d^2}{dz^2} + [2- z (a_+ -  a_- -1)]\frac{d}{dz} -a_+ a_- \right] G(z) = 0.
\end{equation}
where we define $a_+$ and $a_-$ as follows:
\begin{equation}
 a_+ = \frac{3}{2} +\sqrt{\frac{9}{4} - m^2},
\end{equation}
\begin{equation}
 a_- = \frac{3}{2} - \sqrt{\frac{9}{4} - m^2}.
\end{equation}
 The propagator for the massive free scalar field  is given by
 \begin{equation}
 G(z) = \frac{1}{16\pi^2}\Gamma(a_+)\Gamma(a_-) F(a_+ , a_- , 2, z).
 \end{equation}
 Here $F(a_+  , a_- , 2, z)$ is the Gauss hypergeometric function.
The normalization constant is
fixed by analyzing the short 
distance behaviour of the two-point function, 
which should resemble flat spacetime case. 

\section{Vector Field Theory on de Sitter Spacetime}
Now we can define vector harmonics on $S^n $ as 
Let $A_a^{Llp\cdots}$ \cite{ql6}-\cite{ql7}. 
The LB operator be also defined to be an operator which acts on vector harmonics as follows, 
\begin{equation}
  -\nabla^2_n A_a^{Llp\cdots} = [L(L+n-1)-1] A_a^{Llp\cdots}.
\end{equation}
Now for four dimensional vector spherical harmonics on $S^4$, the action of  LB operator  will be given by 
\begin{equation}
 -\nabla^2 A^a_{Llpm} = [L(L + 3) - 1] A^a_{Llpm}, 
\end{equation}
here again we denote $-\nabla_4^2$ as $-\nabla^2$.
Furthermore, the divergence of vector spherical harmonics vanishes. Now we can write  
\begin{equation}
 \nabla_a A^a_{Llpm}=0.
\end{equation}
There are two kinds of solutions that these equations satisfy. 
They will be represented by  $A^m_{Llpm}$,  where $m = 1, 2$. 
So, we can write the vector spherical harmonics on $S^4$ as 
\begin{eqnarray}
 A^1_\chi &=& 0 , \\ A^1_i& =& n_1 P_{L+1} Y_i, \\
A^0_\chi &=& n_2 (\sin\chi)^{-2} P_{L+1}Y, \\ A^0_i &=& n_2  \frac{1}{\ell_0\ell_2} D_1 P_{L+1}\nabla_i Y. 
\end{eqnarray}
here $n_1$ and $n_2$ are normalization constants. They  satisfy 
\begin{equation}
 \int d^4 x \sqrt{g} g^{ab} A_a^{L} A_b^{*L'} = \delta_{LL'}.
\end{equation}

The massive vector field equation of motion on de Sitter spacetime is given by
\begin{equation}
 \nabla_b(\nabla^b A^a - \nabla^a A^b) - m^2A^a = 0.
\end{equation}
This can be written as 
\begin{equation}
 (\nabla^2 - 3 - m^2 ) A^a =0. 
\end{equation}
Thus after analytical continuation from $S^4$ to de Sitter,  we have 
\begin{equation}
 L (L+3) - m^2 -4 =0. 
\end{equation}
Thus we get 
\begin{equation}
L = b_{\pm} = - \frac{3}{2} \pm \sqrt{\frac{24}{4}- m^2}, 
\end{equation}
Here the solution to this equation is of the following form
\begin{equation}
 Q_{ab'}(z) = \alpha^V(z) g_{ab'} +\beta^V(z) n_a n_{b'},
\end{equation}
where
\begin{equation}
 \alpha^V(z) = \left[ \frac{-2z(1-z)}{3}\frac{d}{dz} + 2z -1\right ] \gamma(z)
\end{equation}
and
\begin{equation}
 \beta^V(z) = \alpha^V(z) - \gamma(z).
\end{equation}
 $\gamma$ is given by
 \begin{equation}
 \left[z(1-z)\frac{d^2}{dz^2} + [2- z( b_+ -  b_- -1)]\frac{d}{dz} -b_+ b_- \right] \gamma(z) = 0.
\end{equation}
Here
\begin{equation}
 b_+ = - \frac{3}{2} + \sqrt{\frac{24}{4}- m^2}, 
\end{equation}
and
\begin{equation}
 b_- = -\frac{3}{2} - \sqrt{\frac{24}{4}- m^2}.
\end{equation}
 Now $\gamma(z)$ will given by
\begin{equation}
 \gamma(z) = \frac{-3 \Gamma(b_+)\Gamma(b_-)}{64\pi^2 m^2}F(b_+, b_- , 3, z).
\end{equation}

\section{Conclusion}
In this paper we analysed quantum field theory on de Sitter spacetime. 
We quantized a vector and a scalar field theory on de Sitter spacetime. 
It may be noted that this formalism can  be applied to anti-de Sitter spacetime. 
We can also try to apply this formalism to gravity. In fact, recently higher spin fields have become 
very important. It might be possible to apply this formalism to higher spin field too. 
It may be interesting to use the techniques of this paper to study massless vector fields. 
However, massless vector fields have a gauge symmetry associated with them. 
They can thus not be directly quantized without fixing a gauge. Thus, we need to fix a gauge. 
This can be done at a quantum level by adding a gauge fixing term and a ghost term 
to the original classical action. This  can be done first on four sphere and then the results can 
be continued to de Sitter spacetime. It may be noted that we might get infrared divergences 
if we try to calculate these ghosts. However, as the ghosts do not couple to the vector fields 
for abelian gauge theories, we need not worry about these divergences. They will not influence the 
final results of calculations and the $S$-matrix can be calculated using the above mentioned method.


\begin{thebibliography}{99}
\bibitem{a}
L. Fabbri, Int. J. Theor. Phys. 50, 3616 ,2011
\bibitem{lq}
T. Kugo and I. Ojima, Nucl. Phys. B144, 234 ,1978
\bibitem{pp}
K. Nishijima and M. Okawa, Prog. Theor. Phys. 60, 272 ,1978
\bibitem{lpq}
N. Nakanishi and I. Ojima, Covariant operator formalism of gauge theories
and quantum gravity - World Sci. Lect. Notes. Phys - ,1990
\bibitem{e5}
A. Lesov, arXiv:0911.0058
\bibitem{d5}
J. MacDonald and D. J. Mullan, Phys. Rev. D 80, 043507 ,2009 
\bibitem{d6}
A. M. Sinev, arXiv:0806.3212
\bibitem{l}
C. Pagliarone, arXiv:hep-ex/0612037 
\bibitem{l1}
M. Faizal,  Class. Quant. Grav. 29, 035007 ,2012
\bibitem{airq}
A. Pakman, JHEP 0306, 053, 2003
\bibitem{ir}
M. Faizal and A. Higuchi,Phys. Rev. D85: 12402, 2012 
\bibitem{ir1a}
K. Izumi and T. Tanaka, Prog. Theor. Phys. 121, 427,  2009 
\bibitem{b}
M. Faizal and A. Higuchi, Phys. Rev. D 78,  067502 , 2008
\bibitem{e}
M. Faizal, J. Phys. A 44, 402001, 2011
\bibitem{10a}
I. A. Batalin and G. A. Vilkovisky, Phys. Lett. B 102,  27 ,1981
\bibitem{24}
I. A. Batalin and G. A. Vilkovisky, Phys. Rev. D 28, 2567 , 1983
\bibitem{11a}
C. Bizdadea and S. O. Saliu, J. Phys. A 31, 8805 , 1998
\bibitem{12a}
C. Bizdadea, I. Negru and S. O. Saliu, Int. J. Mod. Phys. A 14, 359 , 1999
\bibitem{f}
M. Faizal, Found. Phys. 41, 270  , 2011
\bibitem{sm}
M. Faizal, Phys. Lett. B 705, 120 , 2011 
\bibitem{smssss}
J. W. Moffat, Phys. Lett. B 506,  193 , 2001
\bibitem{sasassas}
 J. W. Moffat, Phys. Lett. B 491, 345 , 2000 
\bibitem{smu}
M. Faizal, Mod. Phys. Lett. A27: 1250075, 2012 
\bibitem{91}S. Ahmad, Comm. in Theo.l Phys,  59, 439 , 2013 
\bibitem{7a}
M. Faizal, Phys. Rev. D 84,  106011 , 2011
\bibitem{z6}M. Faizal and D. J. Smith, Phys. Rev. D85: 105007, 2012 
\bibitem{cfjcbljkbhgf} V. Mader. M. Schaden, D. Zwanziger and R. Alkofer,  arXiv:1309.0497
\bibitem{dcvasd}M. Faizal, JHEP 1204: 017, 2012 
\bibitem{z61}
A. Gustavsson,  arXiv:1203.5883
 \bibitem{z7}M. Faizal, Comm. Theor. Phys. 57, 637 , 2012
\bibitem{z7a}
M. S. Bianchi, M. Leoni and S. Penati,  arXiv:1112.3649 
\bibitem{z9}M. Faizal, Europhys. Lett. 98: 31003, 2012
\bibitem{fcxf3ffrcwcffcre}
M. Marino and P. Putrov, arXiv:1110.4066
\bibitem{zpa}
K. Okuyama, arXiv:1110.3555
 \bibitem{za}M. Faizal, JHEP. 1204, 017  , 2012
\bibitem{zax}
A. Belhaj,  arXiv:1107.2295 
\bibitem{cds}M. Faizal, arXiv:1303.5477 
\bibitem{dcasd}D. Zwanziger, AIPConf.Proc.892:121-127, 2007 
\bibitem{cdsb}M. Faizal, Mod. Phys. Lett. A28: 1350034, 2013
\bibitem{dcfgbdfasd}D. Zwanziger, Phys. Rev. D76: 125014, 2007 
\bibitem{cdsvn}M. Faizal, Int. J. Mod. Phys. A28: 1350012, 2013 
\bibitem{dcgdasd}M. Golterman, L. Zimmerman, Phys.Rev. D71,  117502 , 2005
\bibitem{cdsbvnb}M. Faizal, JHEP. 1301: 156, 2013 
\bibitem{dcagdbfbsd}D. Polyakov,  Phys.Lett. B611,  173  , 2005
\bibitem{cdsnx}M. Faizal, Europhys. Lett. 103: 21003, 2013 
\bibitem{dcasghghd}A. Imaanpur, JHEP 0503 , 030 , 2005
\bibitem{cdnbgxs}M. Faizal, Nucl. Phys. B. 869: 598, 2013 
\bibitem{dcdghasd}Jen-Chi Lee, Eur.Phys.J.C1:739-741,1998 
\bibitem{cdnxcs}M. Faizal,Phys. Rev. D87: 025019, 2013 
\bibitem{dcasfgssd}A. Kapustin, Y. Li, Anton Kapustin, Adv.Theor.Math.Phys. 9 , 559, 2005 
\bibitem{cdsnbv}M. Faizal, Int. J. Theor. Phys. 52: 392, 2013 
\bibitem{dcassfd}Chuan-Tsung Chan, Jen-Chi Lee, Yi Yang, Phys.Rev. D71  086005 , 2005
\bibitem{cdnxcbs}M. Faizal,Class. Quant. Grav. 29: 215009, 2012 
\bibitem{dcasgfsd}R. P. Malik, arXiv:hep-th/0412333 
\bibitem{vdfsv}M. Faizal, Comm. Theor. Phys. 58: 704, 2012 
\bibitem{dcasfsgd}N. Boulanger, J.Math.Phys. 46 , 053508  , 2005 
\bibitem{fdvszdf}M. Faizal, Mod. Phys. Lett. A27: 1250147, 2012 
\bibitem{8a}
W. H. Huang, arXiv:1107.2030 
\bibitem{m}
M. Faizal and M. Khan, Eur. Phys. J. C 71, 1603 ,2011 
\bibitem{nu1}
J. T. Liu and Z. Zhao, arXiv:1108.5179 
\bibitem{nu2}
M. Fontanini and M. Trodden, Phys. Rev. D 83, 103518 , 2011
\bibitem{m1}
V. O. Rivelles, Phys. Lett. B 577, 137 , 2003
\bibitem{wd}
B. S. DeWitt,   Phys. Rev. 160,  1113 , 1967
\bibitem{1234}M. Faizal, J.Exp.Theor.Phys. 114 ,  400, 2012, arXiv:gr-qc/0602094
\bibitem{12345}
Y. Ohkuwa,  Int. J. Mod. Phys. A 13, 4091  , 1998 
\bibitem{1245}M. Faizal,  Mod. Phys. Lett. A 27,   1250007 , 2012
\bibitem{sdcvcad}I. T. Durham,  arXiv:1307.3691 
\bibitem{dfvs}M. Faizal, arXiv:1304.0259 
\bibitem{fcvsdfccvcsa}V. Bonzom, Phys.Rev.D84:024009, 2011
\bibitem{fvszdv}M. Faizal, arXiv:1303.5478 
\bibitem{asdcvdS}Ru-Nan Huang, arXiv:1304.5309 
\bibitem{ddsadfa}R. M. Wald, 
Quantum Field Theory in Curved Spacetime and Black Hole Thermodynamics- University of Chicago Press- 1994
\bibitem{wd1}M. Faizal, arXiv:1301.0224 
\bibitem{ql1}S. W. Hawking and G. F. R. Ellis, The large scale structure of spacetime, 
Cambridge University Press, Cambridge, 1973
\bibitem{fvwcfdvc}R. M. Wald,
Quantum field theory in curved spacetime and black hole thermodynamics, 
University of Chicago Press, Chicago, 1994
\bibitem{fvadfs}N. Katsumi,  Hokk.  Math. J. 11,  253, 1982
\bibitem{ql0}W.  de Sitter,    Proc. Kon. Ned. Acad. Wet. 20,  229, 1917
\bibitem{abcf1}A. G. Riess, A. V. Filippenko, P. Challis, et al, AJ. 116, 1009, 1998
\bibitem{abcf2}S. Perlmutter, G. Aldering, G. Goldhaber, et al, APJ 517, 565, 1999
\bibitem{asymp}M. N. Celerier, New Advances in Physics 1, 29, 2007
\bibitem{ds}C. G.  Huanga,   Y. Tian, X.  Wu and H. Y Guo, Front. Phys. China. 3, 191, 2008
\bibitem{ds1} B. Vakili, N. Khosravi and H. R. Sepangi, Int. J. Mod. Phys. D18, 159, 2009
\bibitem{ds2}M. N. Chernodub, L. Faddeev and A. J. Niemi, JHEP 0812, 014, 2008
\bibitem{ds4} L.  Covi, M.  G. Reino, C.  Gross, J.  Louis, G. A. Palma and C.  A. Scrucca, 
JHEP 0806, 057, 2008
\bibitem{ds5}R. Aros, C. Martinez, R. Troncoso, and J. Zanelli,
 JHEP 05,
 020, 2002
\bibitem{ds6}R.  Aros, Phys. Rev. D77, 104013, 2008
\bibitem{ds7}E. M. Morales, J. Math. Phys. 52, 102304, 2011 
\bibitem{ql2}A. Higuchi and Y. C. Lee,   Phys. Rev. D 78, 084031, 2008
\bibitem{scdcxsd}A. Higuchi and S .S. Kouris,  Class. Quant. Grav. 17,   3077, 2000
\bibitem{csdfcewdc} Class. Quant. Grav. 18,  4961 2001
\bibitem{ql3}A. Higuchi and S. S. Kouris, Class. Quant. Grav. 18,  4317, 2001
\bibitem{ql4} R. Szmytkowski,  J. Math. Chem. 49,  1436, 2011 
\bibitem{ql5} H. S. Cohl,  SIGMA 7, 050,  2011
\bibitem{ql6}A. Higuchi, J. Math. Phys. 28, 1553, 1987
\bibitem{rtfevrfg}N. Vilenkin, Am. Math. Soc. Transl.  22, 1968
\bibitem{vsfvsfs}R. G. Barrera, G. A. Estevez and J. Giraldo, Eur. J. Phys. 6,  287, 1985
\bibitem{fwfdvcw}B. Carrascal, G.A. Estevez, P. Lee and V. Lorenzo,  Eur. J. Phys. 12, 184, 1991
\bibitem{vcwfwv}E. L. Hill,  Am. J. Phys. 22, 211, 1954
\bibitem{vclwfwv}E. J. Weinberg,  Phys. Rev. D. 49, 1086, 1994
\bibitem{vcwlkfwv}P. M. Morse and H. Feshbach, Methods of Theoretical Physics, 
Part II,  McGraw-Hill,   New York, 1953
\bibitem{ql7}  C. Frye and C. J. Efthimiou,   arXiv:1205.3548 
\end{thebibliography}
\end{document}